\documentclass[12pt,journal,draftcls,letterpaper,onecolumn]{IEEEtran}

\usepackage{a4,epic,eepic,graphicx}

\usepackage{booktabs}
 \usepackage{amssymb}
\usepackage{amsmath}
\usepackage{psfrag}

\def\N{{\rm I\kern-.5ex N}}
\def\Z{{\sf \vrule height 1.55ex depth-1.2ex width.03em\kern-.11em Z%
        \kern-.9ex Z\kern-.11em\vrule height 0.3ex depth0ex
width.03em}}
\def\Q{{\rm\kern.2ex\vrule height1.55ex depth-.05ex width.03em\kern-
.7ex Q}}
\def\R{{\rm I\kern-.5ex R}}
\def\C{{\rm\kern.3ex\vrule height1.55ex depth-.05ex width.03em\kern-
.7ex C}}

\newif\ifcomment\commentfalse
\def\commentON{\commenttrue}

\long\outer\def\bc#1\ec{{\ifcomment \sloppy  $[${\bf Leen suggests}]
{{#1}} \textbf{[end]} \fi }}

\long\outer\def\bsc#1\esc{{\ifcomment \sloppy  $[${\bf Steven suggests}]
{{#1}} \textbf{[end]} \fi }}

\long\outer\def\bso#1\eso{{\ifcomment \sloppy  $[${\bf Steven observes}]
{{#1}} \textbf{[end]} \fi }}

\long\outer\def\br#1\er{{\ifcomment \sloppy  $[${\bf suggest remove}]
{{#1}} \textbf{[end]} \fi }}

\long\outer\def\bo#1\eo{{\ifcomment \sloppy  $[${\bf instead of}]
{\textit{#1}} \textbf{[end]}  \fi }}

\long\outer\def\BC#1\EC{{\ifcomment \sloppy \par \#  \dotfill
{\textsc{#1}} \dotfill \# \par \fi }}

\commentON

\begin{document}

\title{Some mathematical refinements concerning error minimization in the genetic code}

\author{Harry Buhrman, Peter T. S. van der Gulik, Steven M. Kelk,\\ Wouter M. Koolen, Leen Stougie\footnote{All authors
are affiliated with the Centrum voor Wiskunde en Informatica (CWI), P.O. Box 94079, NL-1090 GB Amsterdam, The Netherlands,
email \{harry.buhrman, Peter.van.der.Gulik, wmkoolen, s.m.kelk, leen.stougie\}@cwi.nl. Leen Stougie
is also affiliated with the Division of Econometrics and Operations Research, Department of Economics and Business Administration,
Vrije Universiteit, De Boelelaan 1105, 1081 HV Amsterdam, The Netherlands, email lstougie@feweb.vu.nl. Part of this research has
been funded by the Dutch BSIK/BRICKS grant, Vici grant 639-023-
302 from the Netherlands Organization for Scientific Research (NWO),
and the CLS project.
}}

\maketitle

\begin{abstract}
The genetic code has been shown to be very error robust compared to randomly
selected codes, but to be significantly less error robust than a certain code found by a heuristic algorithm. We formulate this
optimisation problem as a Quadratic Assignment Problem and thus verify that the code found by the heuristic is
the global optimum. We also argue that it is strongly misleading to compare the genetic code only with codes sampled from the fixed block
model, because the real code space is orders of magnitude larger. We thus enlarge the space from which random codes can be sampled from
approximately $2.433 \times 10^{18}$ codes to approximately $5.908 \times 10^{45}$ codes. We do this by leaving the fixed block model, and 
using the wobble
rules to formulate the characteristics acceptable for a genetic code. By relaxing more constraints three larger spaces are also
constructed. Using a modified error function, the genetic code is found to be more error robust compared to a background of randomly
generated codes with increasing space size. We point out that these results do not necessarily imply that the code was optimized during
evolution for error minimization, but that other mechanisms could explain this error robustness.
\end{abstract}


\section{Background}
\label{sec:background}

\begin{table}[h]
\caption{The standard genetic code. Assignment of the 64 possible codons to amino acids or stop signals,
with polar requirement of the amino acids indicated in brackets.}
\centering
  \begin{tabular}{ | l | l | l | l | }
    \hline
UUU Phe (5.0) & UCU Ser (7.5) & UAU Tyr (5.4) & UGU Cys (4.8) \\
UUC Phe (5.0) & UCC Ser (7.5) & UAC Tyr (5.4) & UGC Cys (4.8) \\
UUA Leu (4.9) & UCA Ser (7.5) & UAA STOP & UGA STOP \\
UUG Leu (4.9) & UCG Ser (7.5) & UAG STOP & UGG Trp (5.2) \\ \hline

CUU Leu (4.9) & CCU Pro (6.6) & CAU His (8.4) & CGU Arg (9.1) \\
CUC Leu (4.9) & CCC Pro (6.6) & CAC His (8.4) & CGC Arg (9.1) \\
CUA Leu (4.9) & CCA Pro (6.6) & CAA Gln (8.6) & CGA Arg (9.1) \\
CUG Leu (4.9) & CCG Pro (6.6) & CAG Gln (8.6) & CGG Arg (9.1) \\ \hline

AUU Ile (4.9) & ACU Thr (6.6) & AAU Asn (10.0) & AGU Ser (7.5) \\
AUC Ile (4.9) & ACC Thr (6.6) & AAC Asn (10.0) & AGC Ser (7.5) \\
AUA Ile (4.9) & ACA Thr (6.6) & AAA Lys (10.1) & AGA Arg (9.1) \\
AUG Met (5.3) & ACG Thr (6.6) & AAG Lys (10.1) & AGG Arg (9.1) \\ \hline

GUU Val (5.6) & GCU Ala (7.0) & GAU Asp (13.0) & GGU Gly (7.9) \\
GUC Val (5.6) & GCC Ala (7.0) & GAC Asp (13.0) & GGC Gly (7.9) \\
GUA Val (5.6) & GCA Ala (7.0) & GAA Glu (12.5) & GGA Gly (7.9) \\
GUG Val (5.6) & GCG Ala (7.0) & GAG Glu (12.5) & GGG Gly (7.9) \\
    \hline
  \end{tabular}
\label{tab:tab1}
\end{table}
\noindent

The genetic code is the set of rules according to which nucleic acid sequences are translated into
amino acid sequences. Although a few small variations on the standard genetic code are known
(especially in mitochondrial systems), this set of rules is essentially the same for all organisms.
The genetic code is therefore one of the most fundamental aspects of biochemistry. The pattern of
codon assignments in the genetic code appears to be organized in some way (Table \ref{tab:tab1}). First, there is
codon similarity for codons encoding the same amino acid. The underlying biochemical reason
\cite{crick66} is (partly) that tRNA molecules often recognize more than one
codon. A second phenomenon is that similar amino acids are often specified by similar codons. One
way to quantify amino acid similarity is to use the values of polar requirement introduced by Woese
et al. \cite{woese66}. According to this measure amino acids with a polar side chain like glutamate
and aspartate have a high value (12.5 and 13.0, respectively), while hydrophobic amino acids like
leucine and valine have a low value (4.9 and 5.6, respectively). An example of similar codons 
coding for similar amino acids is asparagine, specified by codons AAU and AAC with a polar
requirement of 10.0 and lysine, specified by AAA and AAG, with a polar requirement of 10.1.
Although one may suspect that similar codons code for similar amino acids may also be present in 
a random grouping \cite{crick68}, Haig and Hurst \cite{haighurst91, haighurst99} showed that this 
is not the case. Random codes do not have this property to the same extent as the standard 
genetic code.\\
\\
Haig and Hurst \cite{haighurst91} generated by computer a large number of alternative
genetic codes, in which the blocks coding for amino acids in the standard genetic code, e.g. the
UCU, UCC, UCA, UCG, AGU, AGC block encoding serine, were kept the same, but their assignment to an
amino acid was randomly redistributed (a procedure generally called ``swapping''). We will refer
to this as the \emph{fixed block model} \cite{goldman93}. Note that the use of the word ``block''
is different from the use in studies such as \cite{novozhilovetal2007, senguptaetal2007}. We use 
the word "block" as in \cite{goldman93} and \cite{freelandhurst98}: in
the sense of the collection of all codons specifying the same amino acid or chain
termination ("STOP" in Table \ref{tab:tab1}). We will call the collection of all codons sharing the same first
and second nucleotide "box". The space of codes which is
created as a result of random code generation under the fixed block model, denoted as Space 0, contains 
exactly $20!$ ($\approx 2.433 \times 10^{18}$) codes. \\
\\
As a measure for the quality of a code the change in polar requirement caused by one 
step point mutations in the codons was proposed. Each codon has nine codons to which it can
mutate in one step: e.g. for the UCU serine codon, these are UCC, UCA, UCG (these three remain
coding for serine in the actual code), UUU (coding for phenylalanine, a 2.5 difference in polar
requirement), UAU (coding for tyrosine, a 2.1 difference), UGU (coding for cysteine, a 2.7 difference), CCU
(coding for proline, a 0.9 difference), ACU (coding for threonine, also a 0.9 difference), and GCU (coding
for alanine, a 0.5 difference). The quality of the code is then measured by averaging over all squared
differences: $MS_0$. In this calculation, Haig and
Hurst \cite{haighurst91} ignored the three ``stop codons'' which are coding for chain termination.
In this way, 263 connections between adjacent codons contribute equally to $MS_0$.\\
\\
To facilitate the mathematical formulation of $MS_0$ we introduce an undirected graph $G = (V,E)$
that has the 61 codons as its vertices and an edge between any two codons if they differ in only
one position, yielding 263 edges. Let $G^{S} = (V^{S}, E^{S} )$ be the graph obtained by adding the
3 stop codons to $V$, yielding 288 edges. A code $F$ maps each codon $c$ to exactly one amino acid $F(c)$.
We denote by $r(F(c))$ the polar requirement of the amino acid that codon $c$
encodes for w.r.t. code $F$. The error function of code $F$ is then given by

\begin{eqnarray*}
MS_0(F)= \frac{1}{263} \sum_{\{ c,c'\} \in E} \left( r(F(c))-r(F(c')) \right) ^2.
\end{eqnarray*}

Using $MS_0$ as a quality measure of a genetic code Haig and Hurst found that only 1 out
of 10,000 random codes performs better, i.e. has a lower $MS_0$, than the standard genetic code \cite{haighurst99}. This shows that in the 
standard genetic code not only identical amino acids are
encoded by similar codons, but also similar amino acids are encoded by similar
codons. Originally, Haig and Hurst \cite{haighurst91, haighurst99} investigated three other characteristics beside
polar requirement (like e.g. the isoelectric point), but the
correspondence between codon assignments and error robustness with
respect to polar requirement was most striking. It may be interesting to find other measures which perform equally well, or 
better. However, the measure has to be independent from the genetic code (this point has been made in connection with the use of values 
derived from replacement mutations known from sequence data). We have to be careful not to artificially create a 
measure that is based on the genetic code itself. To keep results comparable to the work of Haig and Hurst, use of polar requirement is 
preferable. \\
\\
The work of Haig and Hurst was soon followed by the work of Goldman \cite{goldman93}, who found
a code using a heuristic method that has a lower $MS_0$ value than any of the codes generated before.
In Section \ref{subsec:goldmanbest} we verify that Goldman's code is in fact the global optimum in the fixed block model.\\
\\
Freeland and Hurst \cite{freelandhurst98} presented four histograms to visualise the particular error
robustness, in the sense of Haig and Hurst \cite{haighurst91}, of the standard genetic code. They
reported that with respect to the $MS_0$ value, 114 codes out of the 1,000,000 random
codes had a lower value than the standard genetic code. They also reported similar results with respect to
the MS measure restricted to point mutations in the first, second and third
codon, respectively denoted by $MS_1$, $MS_2$ and $MS_3$. To define them we
partition the edge set $E$ in the graph representation $G = (V, E )$ of the adjacency
structure of codons, depending on the position in which two adjacent codons differ: $E_1$
is the set of edges between two codons that differ only in the first position, $E_2$ the set
of edges between two codons that differ only in the second position, and $E_3$ the set of
edges between two codons that differ only in the third position. Clearly these sets are
disjoint and $E = E_1 \cup E_2 \cup E_3$. Then for $p = 1,2,3,$
\begin{eqnarray*}
MS_p(F)= \frac{1}{|E_p|} \sum_{\{ c,c'\} \in E_p} \left( r(F(c))-r(F(c')) \right) ^2,
\end{eqnarray*}
where $|X|$ denotes the cardinality of $X$ i.e. the number of elements in $X$. In fact, $|E_1|=87$,
$|E_2|=88$ and $|E_3|=88$. The results of Freeland and Hurst show that there is not much error robustness for mutations in the middle position of the codon;
the third position, however, is extremely robust against changes in polar requirement. \\
\\
Subsequent research following this approach has concentrated on nuancing the error function
\cite{freelandhurst98, ardell98, freelandetal2000, gilisetal2001, higgs2009} or taking a parameter different
from polar requirement as an amino acid characteristic \cite{ardell98, freelandetal2000,
gilisetal2001, higgs2009}. The common theme in most of these approaches is the code space from which random
alternative codes are generated; in \cite{freelandetal2000} this space is referred to as
``possible code space'' and we denote this space as Space 0. Remarkably, known genetic code variations lie \emph{outside}
Space 0. In code variations certain individual codons are \emph{reallocated} from
one block to another. The fixed block structure of the standard genetic code is thereby replaced by
an alternative, slightly different, fixed block structure. In Section \ref{subsec:enlarging} we
construct four progressively larger code spaces (denoted Space 1, Space 2, Space 3 and Space 4), which encompass successively more known genetic
code variations next to the standard genetic code. To be able to compare the genetic code with respect to 
alternative codes sampled randomly from Spaces 1 and 2, we nuance
the MS measure such as to accommodate values of polar requirement for stop codons. In this paper, we aim at refining several points
in the seminal work by  Haig and Hurst \cite{haighurst91, haighurst99}, Goldman \cite{goldman93} and Freeland and Hurst \cite{freelandetal2000}. Apart 
from determining the global minimum, the refinements concern the code space
structure and the kind of conclusions assumed to be possible to draw based on the research. We do 
not intend to change the characteristic taken to represent the amino acid (which is polar
requirement in the work of Haig and Hurst \cite{haighurst91, haighurst99} and Goldman \cite{goldman93}) or to weigh the three positions
of the codons differently in the error function (as is done in the second
part of \cite{freelandhurst98} and most subsequent work). We only intend to enlarge the space from which
random codes are sampled, and find out how they relate to \cite{freelandhurst98}.

\section{Results}
\label{sec:results}

\subsection{Goldman's best solution is the global minimum}
\label{subsec:goldmanbest}

\noindent Goldman \cite{goldman93} applied a heuristic algorithm for finding the best code under the
\emph{fixed block model}. The best solution he found had an $MS_0$ value of 3.489, which was well
below the value of 5.194 reported by Haig and Hurst \cite{haighurst91, haighurst99} for the standard
genetic code. A heuristic does not guarantee that the code found is optimal. We designed an exact
method for finding the optimal code by formulating the minimization problem as a Quadratic
Assignment Problem (QAP) \cite{Cela98} and solved it using the exact QAP-solver QAPBB \cite{qapbb}. An intuitive formulation of QAP is as follows. We are given two sets of objects $V_1$ and $V_2$ of equal size. We are to match each object from $V_1$ to exactly one object from $V_2$ such that all objects of $V_2$ are matched as well; as a result we get a perfect matching (pairing) of the objects of $V_1$ and $V_2$. In the ordinary (linear) assignment problem, there is a cost for assigning object $i$ from $V_1$ to object $k$ from $V_2$ and we wish to find the assignment that minimizes total cost. In QAP the cost is dependent on pairs of assignments: there is a cost for assigning object $i$ from $V_1$ to object $k$ from $V_2$ {\em and} object $j$ from $V_1$ to object $\ell$ from $V_2$. Again we wish to minimize the total assignment cost. \\

\noindent If we consider the set of objects $V_1$ to be the 20 blocks in Table \ref{tab:tab1}, and the set of objects $V_2$ to
be the 20 amino acids, then we can model the minimization of $MS_0$ by letting the cost of assigning one amino acid to one block and another amino acid to another block be given by the difference of their polar coordinates times the number of point mutations between the two blocks. In Section~\ref{subsec:qap} we define this problem formally as a 0-1 integer program with quadratic objective and linear constraints. \\
\\
QAP is an NP-hard problem, meaning that it is probably hard to solve \cite{gareyjohnson}. However,
small instances of QAP can be solved effectively using an exhaustive enumeration technique known as
\emph{branch and bound} \cite{qapsurvey}. This searches (implicitly) through the entire space of solutions, keeping note of the best
solution found so far, and ignoring parts of the solution space that could not possibly lead to a better solution.
Even with branch and bound it is in general not feasible to use the QAP model for finding a code
with minimum $MS_0$ value in any reasonable time when we leave the \emph{fixed block model}.
However, we could find the global minimum $MS_0$ value in Space 0. We found the same
solution as Goldman, certifying that his solution was in fact the optimal one.

\subsection{Incorporating stop codons}
\label{subsec:stopcodons}

Leaving the fixed block model required us to nuance the MS measure and attach a value of polar requirement
to the stop signal. Chain termination is produced by Release Factors (RFs), which are proteins, and therefore most
probably later elements of the coding system than tRNAs. This is an argument which can also be found in e.g. \cite{higgs2009} (``... I do not
want to assume that there were stop codons in the current positions from the beginning, because it is more likely that stop codons were a late
addition to the code, after the main layout of most of the codons was already established''). Genetic codes lacking stop codons are not
impossible. During the evolutionary development of the genetic code, mRNAs could have been short, and the last sense codon of a message
could have been the end of the mRNA. After attaching the last amino acid of the polypeptide, the primordial ribosome could move further along the
mRNA, and both the polypeptide and the mRNA could lose the association with the ribosome, as the tape leaving the tape recorder in the classical
analogy. The more sophisticated mechanism with Release Factors could have evolved later, to make things run more smoothly. When this is the scenario 
of evolution of chain termination we follow, we want the stop codons to have the smallest influence on our calculations possible.\\
\\
\emph{How to assign values to stop codons?}\\
\\
There are at least four possible ways to deal with the stop codons. In the work described in Section \ref{sec:background} the stop codons were ignored and no value 
was assigned to them. A second way to deal with stop codons is to assign a fixed value to a stop codon. A third way would be to assign a fixed value 
to the mutation to a stop codon, which would be the same for all amino acids. The last way to deal with the problem would be to mimic the natural 
process of suppression.\\

\subsubsection{Assigning no value to stop codons}

Ignoring stop codons in the calculation as has been done until now \cite{haighurst91, goldman93, freelandhurst98} is not the way in which
their influence is the smallest possible. This is because they eliminate a lot of the edges from $G^S$. For the UCA serine codon, in the previous
treatment only the edges to UCU, UCC, UCG, UUA, CCA, ACA and GCA take part in the calculation. The edges to UAA and UGA are
ignored, which means in fact that they behave towards serine as if those codons were encoding serine. Due to this effect, the four alanine
codons have a stronger influence on the calculation than the four glycine codons. Thus ignoring stop codons artificially favors certain amino 
acids. This effect will even become more pronounced when we enlarge the space of possible codes. For example, if we allow codes to have 
as many as four stop codons (like our mitochondrial code), or to have stop codons in
unusual places (like the UUA and UUG stop codons of the mitochondria of \emph{Pycnococcus provasolii} \cite{turmelletal2010}).  \\

\subsubsection{Assigning a fixed value to a node (i.e. give the stop codon a fixed value)}

If we were to reason that a mutation to a stop codon would lead to truncation of messages, we might be inclined to attach
a very large value to a stop codon (because truncated proteins would be non-functional and the mutation therefore lethal). To model ``lethal'', we
could use the value ``infinity''. This makes our calculation useless. We could also attach a polar requirement of 1,000,000 to a stop codon. In this 
case the stop codons are going to dominate the calculation and this is exactly what we didn't want to begin with. \\

\subsubsection{Assigning a fixed value to an edge (i.e. give the mutation to a stop codon a fixed value)}

There is another way to model the concept that a mutation to a stop codon is worse than a mutation to a sense codon. One could assign a fixed penalty to 
a mutation to "stop", no matter which amino acid is mutated to stop. One relatively large value which could be given as a penalty is the difference in 
polar requirement between the two most dissimilar amino acids. The disadvantage of this approach is again the domination of the calculation by the stop 
mutations. Although less dominating than the very high fixed values suggested for the stop nodes, this approach still has the stop codons dominating the 
calculation, and possibly obscuring the phenomenon we want to see. \\

\subsubsection{The suppression approach}
 
\begin{figure}[h]
\centering
\vspace{0.2cm}
\tiny
\psfrag{MS0}{\raisebox{-0.15cm}{$MS^S_0$}}
\psfrag{MS1}{\raisebox{-0.15cm}{$MS^S_1$}}
\psfrag{MS2}{\raisebox{-0.15cm}{$MS^S_2$}}
\psfrag{MS3}{\raisebox{-0.15cm}{$MS^S_3$}}
\includegraphics[width=1.0\textwidth]{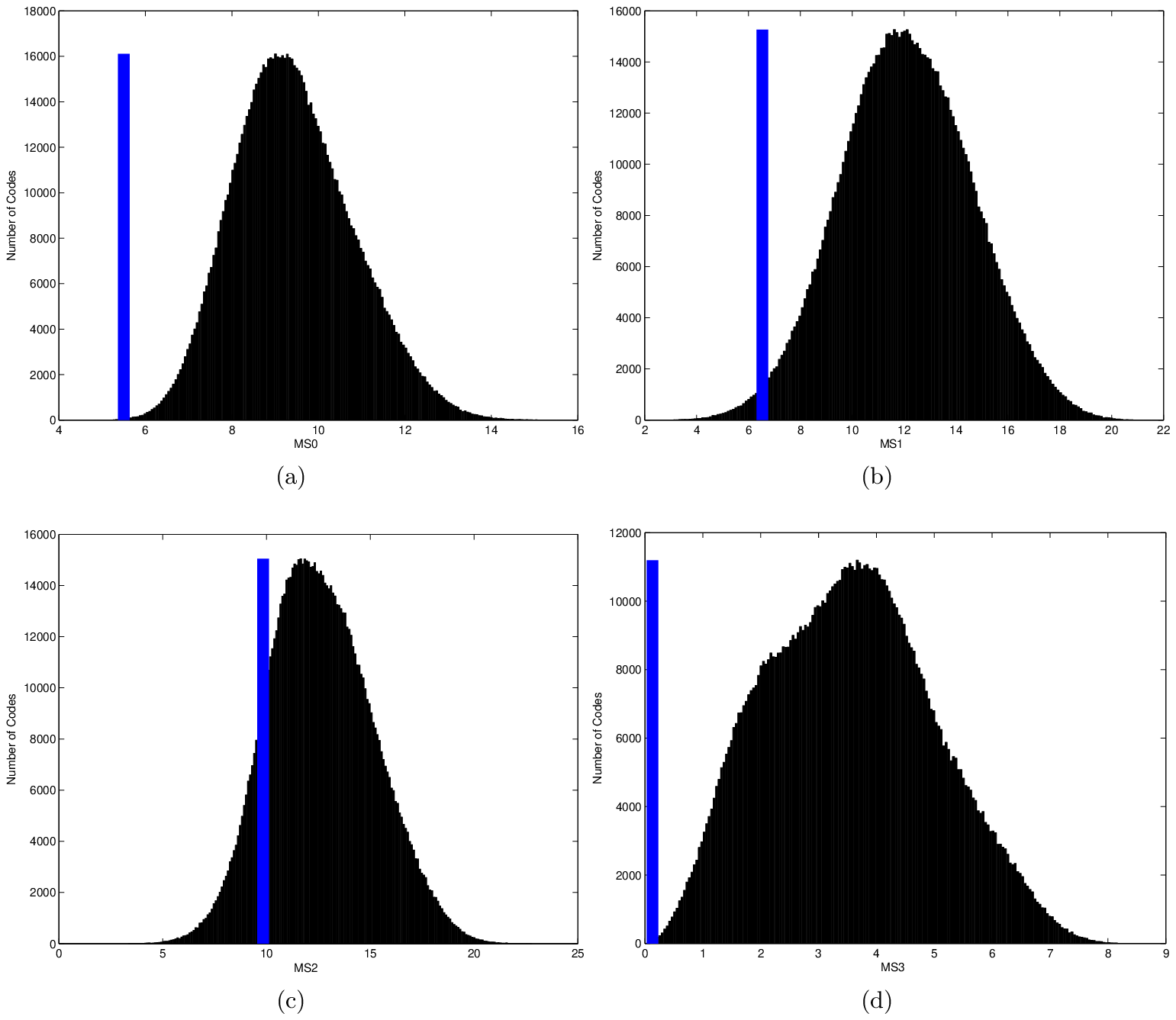}
\caption{\tiny Histograms for the MS values obtained from codes randomly sampled from Space 0. MS value of the standard genetic code indicated by the blue bar. $10^6$ samples. The
MS measure was slightly modified in comparison to earlier work. The modification does not change the basic characteristics found there. (a) $MS^S_0$ (b) $MS^S_1$ (c) $MS^S_2$ (d)
$MS^S_3$.}
\label{fig:samplesfinal1}
\end{figure}

What would happen if there is a mRNA with a codon which does not have a tRNA? In such a case, one possibility is that decoding is performed by the tRNA which, among 
the tRNA repertoire present in the system in consideration, is the most similar to the one which would be needed to decode the codon regularly. This phenomenon is called
``suppression'' in molecular biology \cite{suppression}. In the living cell, the cognate tRNA or RF competes with several different potential suppressor tRNAs
for decoding a codon \cite{kramerfarabaugh2007}. By using in the calculation the value which would be there in case of the most probable suppression, a value is 
attached to a stop codon which results in a relatively small influence of the stop codons in the calculation. The most probable suppression for a stop codon ending on A, is by 
the tRNA which recognizes the sense codon ending on G from the same box. This is reflected by genetic code variants: apparently suppressing tRNAs often evolve towards full 
recognition. We can illustrate this with the UGA codon, which can be found in the top right-hand corner of Table \ref{tab:tab1}. Because the most 
probable suppression for UGA is by the tRNA which normally reads UGG as tryptophan, genetic code variants in which both UGA and UGG 
encode tryptophan evolved multiple times. Although there exists an organism in which UGA is encoding cysteine, the more frequent reassignment for UGA is to 
tryptophan. The same phenomenon is found for AUA, which can be found towards the bottom left-hand side of Table \ref{tab:tab1}. 
AUA has been reassigned several times to methionine. Suppression of AUA codons in protein coding sequences by the tRNA which is normally reading the AUG codons has 
apparently been followed by the evolution of full recognition of the AUA codon by this tRNA. Assigning to a stop codon ending on a purine (A or G)
the value of polar requirement of the amino acid specified by the other purine-ending codon in the box is therefore a possible way to deal with 
stop codons. This obviously can not be done when both purine-ending codons in a box are stop codons. Genetic code variants suggest an approach also in this case. In
bilaterian mitochondria the tRNA which decodes AGA and AGG (recall Table \ref{tab:tab1}, the AGA and AGG codons can be found towards the bottom right-hand side) as arginine 
in the standard code is not present. The tRNA which decodes AGU and AGC as serine usually takes over the function of decoding AGA and AGG by reading them as 
serine \cite{senguptaetal2007}. This suggests the approach: if in 
one box both purine-ending codons are stop codons, the value of polar requirement of the amino acid specified by the codons ending on a pyrimidine (U or C) in 
that box can be assigned to them. This is always a single amino acid because the two pyrimidine-ending codons in the same box always code for the same amino acid. 
Until now, no genetic code variants are discovered with pyrimidine-ending stop codons, so our approach is to develop only a way to deal with stop codons ending on purines. \\
\\
\emph{How to modify the MS measure?}\\
\\
By treating the stop codons as sense codons according to the suppression approach, we simplified the MS measure. In the notation introduced before,

\begin{eqnarray*}
MS^S_0(F)= \frac{1}{|E^S|} \sum_{\{ c,c'\} \in E^S} \left( r(F(c))-r(F(c')) \right) ^2,
\end{eqnarray*}
and similarly w.r.t. the three positions $p=1,2,3$ of the codons
\begin{eqnarray*}
MS^S_p(F)= \frac{1}{|E^S_p|} \sum_{\{ c,c'\} \in E^S_p} \left( r(F(c))-r(F(c')) \right) ^2.
\end{eqnarray*}
In this way, all 64 codons contribute equally to the error measure. Note that
$|E^S|= 288$ and that $|E^S_1| = |E^S_2| = |E^S_3| = 96$. It should be realized that by using $MS^S_0$ or $MS^S_p$ we do
not necessarily start working in a space larger than Space 0. We can use $MS^S_0$ and $MS^S_p$ when we generate random
codes from Space 1 or Space 2 (see Section \ref{subsec:enlarging}) but we can also use $MS^S_0$ and $MS^S_p$
when we generate random codes from Space 0. \\
\\
We investigate how the new measure reflects the nature of Space 0 when used as a background to study the standard genetic code (Table
\ref{tab:tab1}). We produce four plots as in \cite{freelandhurst98}. The plots (Figure
\ref{fig:samplesfinal1}) have the same general shape as the four plots in \cite{freelandhurst98}.
In particular, the prominent shoulder at the left side is present in both the $MS_3^{S}$ (Figure
\ref{fig:samplesfinal1}.(d)) and the $MS_3$ \cite{freelandhurst98} frequency distributions.
The spikes present in the plots in \cite{freelandhurst98} are not present.
They are an artifact of rounding errors in both the data and the bin borders of the
histograms. The combination of MS values rounded to two digits after the decimal point
and bin border values which are repetitive binary fractions rounded by the histogram software,
are probably the source of the spikes in \cite{freelandhurst98}.\\

\noindent The global minimum code in Space 0 for the $MS_0^{S}$ measure was also found using the
quadratic assignment approach described in Section \ref{subsec:goldmanbest}. We calculated the average 
of both $MS_0$ and $MS_0^S$ of 1,000,000 randomly generated codes as well as the global minimum in 
Space 0 with respect to both measures (Table \ref{tab:tab2}). Clearly, both measures give similar results. 
We also studied the proportions of random codes better than the standard genetic code with the $MS_0^{S}$ measure. 
Out of 1,000,000 random codes 156 codes had a lower $MS_0^{S}$-value
than the standard genetic code, resulting in a proportion $P_0^{S}$ of 0.000156. This was also
investigated for $p = 1,2$ and $3$ (Table \ref{tab:tab3}). Again the $MS$ and $MS^{S}$ measures give
similar results (as can be detected also from the plots of Figure \ref{fig:samplesfinal1}).\\
\\
We conclude that it is acceptable to replace $MS$ by $MS^S$ to study the character of the standard
genetic code compared to randomly generated ones. $MS^S$ gives the same results in
all essential aspects, and can be used to investigate larger spaces and spaces with different
codons used as chain termination signal.

\begin{table}[h]
\caption{Comparison of $MS_0$ and $MS^{S}_0$. Values were calculated for $10^6$ randomly sampled codes from Space 0. The averages
and variance are shown; $MS_0$ is taken from \cite{freelandhurst98}.}
\centering
\begin{tabular}{ | l | l | l | }
\hline
 &  $MS_0$ & $MS^{S}_0$ \\
\hline
Mean of random codes & 9.41 $\pm$ 1.51 & 9.43 $\pm$ 1.89\\
\hline
Standard genetic code (rounded) & 5.194 & 5.501\\
\hline
Global minimum code (rounded) & 3.489 & 3.946\\
\hline
\end{tabular}
\label{tab:tab2}
\end{table}

\begin{table}[h]
\caption{Comparison of proportions of ``better codes'' for $MS$ and $MS^S$.}
\centering
\begin{tabular}{ | l | l | }
\hline
$MS$ & $MS^S$ \\
\hline
$P_0 = 0.000114$ & $P_0^{S} = 0.000156$ \\
\hline
$P_1 = 0.002964$ & $P_1^{S} = 0.012369$ \\
\hline
$P_2 = 0.221633$ & $P_2^{S} = 0.129075$ \\
\hline
$P_3 = 0.000088$ & $P_3^{S} = 0.000078$ \\
\hline
\end{tabular}
\label{tab:tab3}
\end{table}

\subsection{Enlarging the ``possible code space''}
\label{subsec:enlarging}

Space 0 has a fixed block structure. It is possible to leave this fixed block structure and generate randomly genetic codes, 
without relaxing all biochemical constraints. In this section we develop a method to enlarge the space from which codes are 
sampled randomly, by specification of allowed subdivision of boxes.\\
\\
Space 0 does not even cover all existing genetic codes: the only existing genetic code present in Space 0 is the standard 
genetic code. By studying code variants general rules with respect to the possible ways to construct a genetic code can be 
found. Using these rules, we enlarge the code space progressively. Genetic code variants are derived from the standard genetic 
code, as can be concluded by studying the codon assignments of close relatives. For mitochondrial code variants this is 
recently described in \cite{senguptaetal2007}. The number of code variants apart from mitochondria is very small and it is 
nowadays believed that they all are derived from the standard code (although this was less clear when the very first variants 
were discovered). Although these variants probably emerged after the standard genetic code, we use the larger spaces because 
they contain possible ways for constructing genetic codes with the system found in living organisms on Earth.\\
\\
In the standard genetic code, the box in the top left-hand corner (see Table \ref{tab:tab1}) shows one of several ways 
in which a box can be subdivided according to the codon-anticodon pairing patterns allowed by the simple wobble rules 
\cite{crick66, bergetal2007}. The codons UUU and UUC are assigned to one amino acid, and the codons UUA and UUG to another. 
Recognition of both pyrimidine-ending codons by one tRNA molecule is the wobbling behavior of G in the first position of the 
anticodon as proposed by Crick \cite{crick66}. Modification of U (in the first position of the anticodon) to thio-U restricts 
the wobbling behavior of the tRNA molecule to recognition of both purine-ending codons \cite{takai2006, numataetal2006}. A second 
pattern of subdivision is presented by a box towards the bottom left-hand side of Table \ref{tab:tab1}. 
In this box AUU, AUC and AUA are assigned to one amino acid and AUG is assigned to another. The existence is known of tRNA 
molecules which recognize all three codons in the top of a box \cite{johanssonetal2008}. Recognition of the G-ending codon 
only, is the wobbling behavior of C in the first position of the anticodon as proposed by Crick \cite{crick66}. Therefore, this 
pattern of subdivision of a box can be understood by the pairing characteristics of tRNA molecules. In eight boxes
of Table \ref{tab:tab1} all four codons are assigned to one amino acid, as in the box in the bottom right-hand 
corner: GGU, GGC, GGA and GGG are assigned to the same amino acid. Recognition of all four codons of a box 
seems to be the wobbling behavior of a tRNA molecule with unmodified U in the first anticodon position \cite{lewin2008}. In 
summary, the wobbling behavior of tRNA molecules allows subdivision of boxes with only sense codons in three ways: no subdivision, 
division in a pyrimidine-ending pair and a purine-ending pair, and division in a set of three codons in the top of a box, and a 
single codon at the bottom. Although extensive modifications of anticodons in contemporary organisms can lead to much more complex 
patterns of wobbling behavior \cite{marckgrosjean2002, takaiyokoyama2003, takai2006, johanssonetal2008}, for the purpose of enlarging 
Space 0 we do not take these aspects of the wobble phenomenon in account. These modifications are produced by proteins, and therefore 
were probably not present during the development of the coding system. To allow the modifications of U to thio-U (enabling the exclusive 
recognition of purine-ending codons) and A to I (enabling the recognition of three codons by one tRNA molecule) is already pushing 
the limit concerning capacities credibly attributable to a very early living system.\\
\\
Further subdivisions of boxes are possible when stop codons are added to the possibilities in a box. Because stop codons ending on 
pyrimidines are not discovered yet, we restrict the possibilities to purine-ending stop codons only. This adds four further ways to 
subdivide a box. The upper two codons assigned to an amino acid, and the lower two codons being stop codons is the first. The upper 
three codons assigned to one amino acid, and the bottom codon being a stop codon is the second. The upper two codons assigned to one 
amino acid, the third codon being a stop codon, and the last codon assigned to an amino acid, but different from the amino acid 
assigned to the upper two codons, is the third possibility. The last possibility again has the third codon being a stop codon, but the three remaining 
codons are assigned to the same amino acid in this case. Taken together with the three possibilities for subdivision with only sense 
codons presented in the previous paragraph, we arrive at seven possible ways to subdivide a box according to the simple wobbling 
behaviour without extensive anticodon modification. This is summarized in Table \ref{tab:tab4}. We generate block structures 
uniformly at random according to the rules described in Table \ref{tab:tab4}, the block structures consist of 21 blocks. \\

\begin{table}[h]
\caption{Possible types of boxes. a = amino acid. b = amino acid, different from a. s = stop.}
\centering
\begin{tabular}{| l | p{11cm} |}
\hline
Box & Meaning \\
\hline
AAAA & All 4 codons recognized by the same tRNA (or by several tRNAs carrying the same amino acid). \\
\hline
AAAB & NNU, NNC, NNA recognized by one tRNA, NNG recognized by another tRNA carrying a different kind of amino acid. \\
\hline
AAAS & NNU, NNC, NNA recognized by a tRNA, NNG by a Release Factor (RF). \\
\hline
AABB & NNU, NNC recognized by one tRNA, NNA, NNG by another tRNA carrying a different kind of amino acid. \\
\hline
AASA & NNU, NNC recognized by one tRNA, NNA by a RF, NNG by another tRNA, but carrying the same amino acid.\\
\hline
AASB & NNU, NNC recognized by one tRNA, NNA by a RF, NNG by another tRNA, carrying a different kind of amino acid. \\
\hline
AASS & NNU, NNC recognized by one tRNA, NNA, NNG recognized by a RF.\\
\hline
\end{tabular}
\label{tab:tab4}
\end{table}
\noindent

In our first extension, the ``stop block'' consists of three stop codons,
as in the standard genetic code. However, their location is free, under the condition that they do
not end in U or C. The number of codons allocated to any amino acid is free, as long as each amino
acid is encoded by at least one codon. In this way we obtain a first enlarged space, Space 1, that
is more realistic than Space 0. Space 1 is, with approximately $5.908 \times 10^{45}$ possible codes, much larger than
Space 0 (with approximately $2.433 \times 10^{18}$ codes).\\
\\
To include most existing genetic code variations, which differ in the number of stop codons, we
enlarged Space 1 to Space 2, by allowing the codes to have $0-4$ stop codons. \\
\\
For completeness, we also define two more spaces but we will not use them in our calculations. In 
some bacteria some codons are not used: neither tRNAs nor release factors to recognize
them (without suppression) are present. To include these code variations too we in addition add 
a new block ``unassigned'' to our block structure, allowing the number of unassigned codons to
range between $0$ and $40$ (Space 3). Every codon is allowed to be unassigned, with the restriction that
codons ending on U or C are either both assigned or both unassigned. Space 3 contains all existing natural 
genetic code variations.\\
\\
Finally (Space 4) we also include codes with fewer or more than 20 amino acids. In many speculations on
the origin of the genetic code, codes with less than 20 amino acids play a role; Jukes suggested such an
evolutionary pathway already in 1966 \cite{jukes}. With the extreme of just one codon in use, the
number of unassigned codons ranges from $0$ to $63$. The size of Space 4 is approximately $1.120
\times 10^{50}$ codes. The sizes of Spaces 0-4 are presented in Table \ref{tab:tab5}. In Section~\ref{subsec:counting} we 
explain the methods behind sampling and counting. The presence of unassigned codons in Spaces 3 and 4 causes 
the function $MS^{S}$ to be ill-defined. Therefore we could not investigate the nature of these spaces, as 
we will do for Spaces 1 and 2. \\
\\

\begin{table}[h]
\caption{Sizes and characteristics of the five progressively larger spaces. Number of codes present in Spaces $0-4$. The block structure of Spaces $1-4$ is free, except for the
constraints imposed by adherence to the Wobble Rules, and the specifications listed under
``Characteristics of codes''.}
\centering
\begin{tabular}{|l|p{7cm}|p{3cm}|}
\hline
Space & Characteristics of codes & Approximate size of space \\
\hline
Space 0 & 21 blocks, 20 amino acids, 3 stop codons, 0 unassigned codons & $2.433 \times 10^{18}$ \\
\hline
Space 1 & 21 blocks, 20 amino acids, 3 stop codons, 0 unassigned codons, free block structure & $5.908 \times 10^{45}$ \\
\hline
Space 2 & 20-21 blocks, 20 amino acids, 0-4 stop codons, 0 unassigned codons, free block structure & $1.932 \times 10^{46}$ \\
\hline
Space 3 & 20-22 blocks, 20 amino acids, 0-4 stop codons, 0-40 unassigned codons, free block structure & $8.635 \times 10^{48}$ \\
\hline
Space 4 & 2-34 blocks, 1-32 amino acids, 0-4 stop codons, 0-63 unassigned codons, free block structure & $1.120 \times 10^{50}$ \\
\hline
\end{tabular}
\label{tab:tab5}
\end{table}

\begin{figure}[h]
\centering
\vspace{0.2cm}
\tiny
\psfrag{MS0}{\raisebox{-0.15cm}{$MS^S_0$}}
\psfrag{MS1}{\raisebox{-0.15cm}{$MS^S_1$}}
\psfrag{MS2}{\raisebox{-0.15cm}{$MS^S_2$}}
\psfrag{MS3}{\raisebox{-0.15cm}{$MS^S_3$}}
\includegraphics[width=1.0\textwidth]{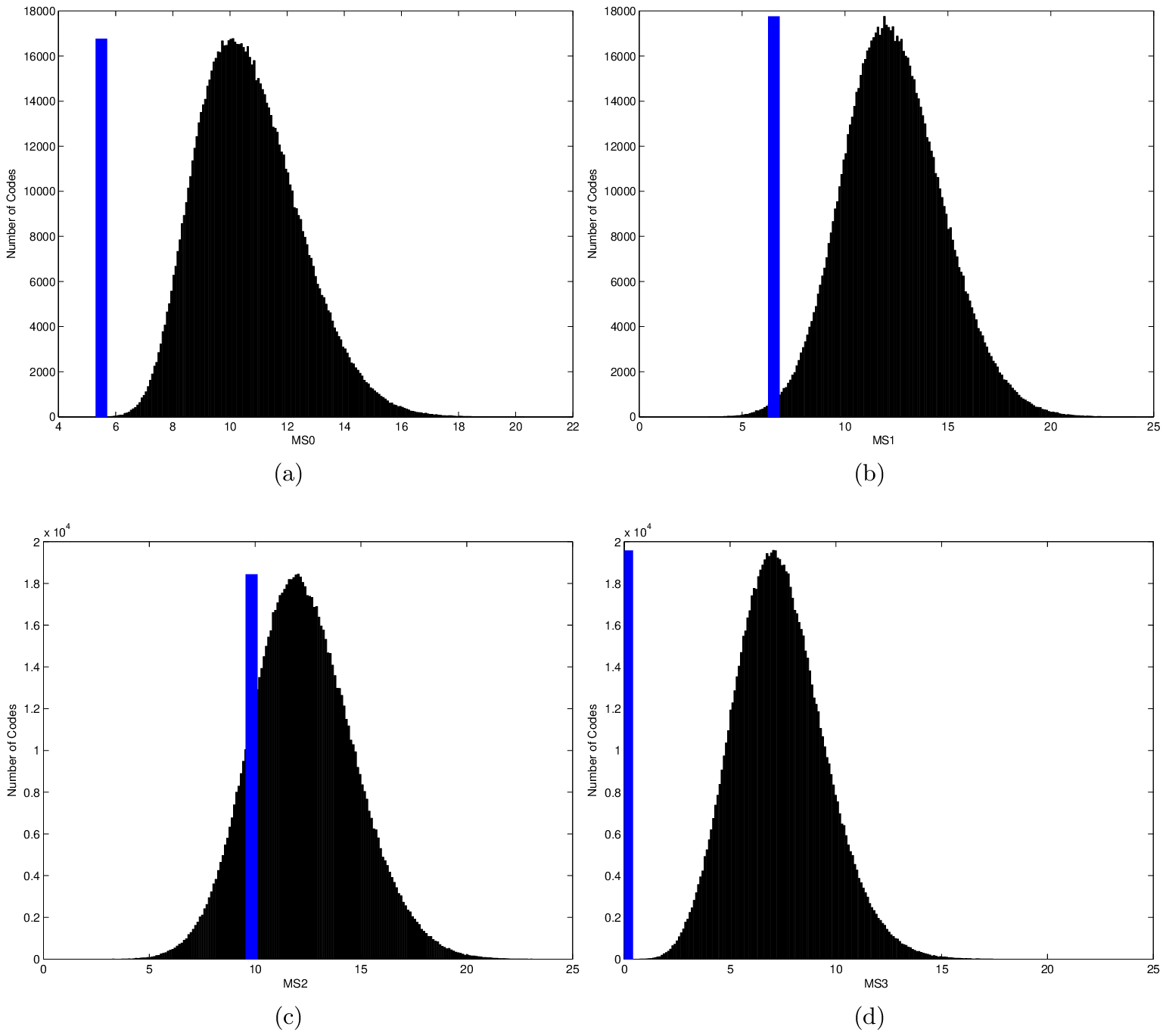}
\caption{\tiny Histograms for the MS values obtained from codes randomly sampled from Space 1. MS value of the standard genetic code indicated by the blue bar. $10^6$
samples. The modified MS measure was used to calculate a MS value because the random redistribution of the three
stop codons made the use of the MS measure from earlier work impossible. The distribution of randomly generated codes is more regular with respect to changes in the third
codon position compared with that distribution resulting from codes sampled from Space 0
(shown in Figure \ref{fig:samplesfinal1}). (a) $MS^S_0$ (b) $MS^S_1$ (c) $MS^S_2$ (d) $MS^S_3$.
}
\label{fig:samplesfinal2}
\end{figure}

\begin{figure}[h]
\centering
\vspace{0.2cm}
\tiny
\psfrag{MS0}{\raisebox{-0.15cm}{$MS^S_0$}}
\psfrag{MS1}{\raisebox{-0.15cm}{$MS^S_1$}}
\psfrag{MS2}{\raisebox{-0.15cm}{$MS^S_2$}}
\psfrag{MS3}{\raisebox{-0.15cm}{$MS^S_3$}}
\includegraphics[width=1.0\textwidth]{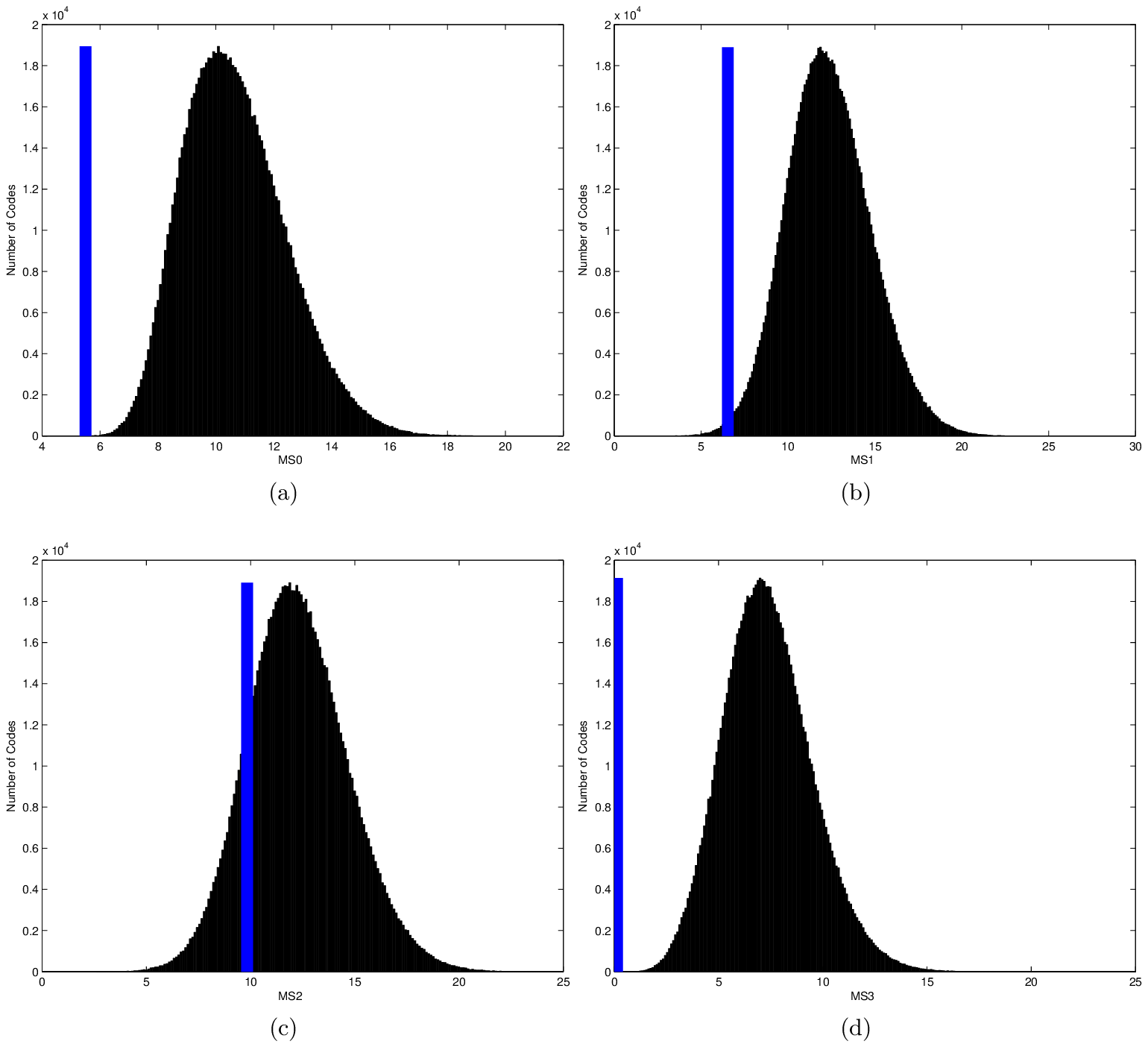}
\caption{\tiny Histograms for the MS values obtained from codes randomly sampled from Space 2.
MS value of the standard genetic code indicated by the blue bar. $10^6$ samples. The modified
MS measure was used to calculate a MS value because random redistribution of $0-4$ stop
codons made the use of the MS measure from earlier work impossible. The behavior
of the distributions is virtually the same as that sampled from Space 1 and shown in the previous
figure. (a) $MS^S_0$ (b) $MS^S_1$ (c) $MS^S_2$ (d) $MS^S_3$.
}
\label{fig:samplesfinal3}
\end{figure}

\noindent Figure \ref{fig:samplesfinal2} shows four plots (as in Figure \ref{fig:samplesfinal1}) of $MS^S$-values, but 
of codes sampled from Space 1 rather than Space 0. We notice the great similarity with the plots in 
Figure \ref{fig:samplesfinal1}. Despite the fact that Space 1 is about $2 \times 10^{27}$ times larger than 
Space 0, the mean $MS_0^{S}$-value is still about 10. The frequency distributions have the same general nature, and
the position of the frequency distribution relative to that of the standard genetic code has not changed. We also 
notice that the prominent shoulder at the left side of the $MS_3^{S}$ frequency distribution in 
Figure \ref{fig:samplesfinal1} has disappeared. We conjecture that the particular block structure of the standard 
genetic code is responsible for this shoulder.\\
\\
Figure \ref{fig:samplesfinal3} shows the same four plots for Space 2. It is hard to find
differences with Figure \ref{fig:samplesfinal2}. The genetic code seems a bit more special against
the background with progressively larger spaces: the number of ``better codes'' found with a
million randomly generated codes decreased from 156 in Space 0, via 7 in Space 1, to just a single one (Table \ref{tab:tab6})
in Space 2.\\
\\

\begin{table}[h]
\caption{Basic descriptive statistics of Space 0, Space 1 and Space 2. From each space $10^6$ codes were randomly sampled.}
\centering
\begin{tabular}{|p{3cm}|l|l|l|}
\hline
Measure & Space 0 & Space 1 & Space 2 \\
\hline
Mean $\pm$ variance & & & \\
\hline
$MS_0^{S}$ & $9.426 \pm 1.89$ & $10.663 \pm 3.13$ & $10.665 \pm 3.12$ \\
\hline
$MS_1^{S}$ & $12.100 \pm 6.37$ & $12.362 \pm 5.88$ & $12.368 \pm 5.86$ \\
\hline
$MS_2^{S}$ & $12.627 \pm 6.33$ & $12.270 \pm 5.79$ & $12.278 \pm 5.79$ \\
\hline
$MS_3^{S}$ & $3.550 \pm 2.09$ & $7.358 \pm 4.51$ & $7.348 \pm 4.49$ \\
\hline
Proportion of better codes found & & & \\
\hline
$P_0^{S}$ & 0.000156 & 0.000007 & 0.000001 \\
\hline
$P_1^{S}$ & 0.012369 & 0.004853 & 0.004864 \\
\hline
$P_2^{S}$ & 0.129075 & 0.151506 & 0.150269 \\
\hline
$P_3^{S}$ & 0.000078 & 0.000000 & 0.000000 \\
\hline
\end{tabular}
\label{tab:tab6}
\end{table}

\section{Discussion}
\label{sec:discussion}

We now compare five published possible scenarios concerning the evolution of the genetic code and show they are not 
inconsistent with low MS values. \\
\\
\subsection{Evolution of the genetic code by selection for error minimization}
\label{subsec:error}

The concept that the codon assignments are a feature of living organisms which protects them against damage to the genetic 
information and which is, as such, specifically selected for by natural selection, was first published by Sonneborn \cite{sonneborn}. 
Woese \cite{woese65} elaborated on this concept by pointing out that it is much more probable that translation errors instead 
of mutations in the genome were the errors against which the system in which the genetic code was developed had to be 
protected. The concept and first experiment of producing computer-generated random codes to compare with the genetic code was published by 
Alff-Steinberger \cite{alff}. This author points out that the differences found depending on the codon position suggest 
translation errors rather than mutations as responsible for determining (in part!) the structure of the code. Haig and Hurst 
\cite{haighurst91} developed the MS measure and were able to generate much more random codes than Alff-Steinberger. They 
again found differences depending on codon position, but left the possibility open, that "... the code acquired its major 
features before the evolution of proteins" \cite{haighurst91}, implying that selection for protection against errors in protein-coding 
messages maybe played no role in the evolution of the genetic code. Freeland and Hurst \cite{freelandhurst98} elaborated on 
the work of Haig and Hurst, and presented the code as "one in a million": "We thus conclude not only that the natural genetic 
code is extremely efficient at minimizing the effects of errors, but also that its structure reflects biases in these errors, 
as might be expected were the code the product of selection" \cite{freelandhurst98}. The extreme version of the "Error Minimization 
Hypothesis" would be that all possible codes were tested by natural selection, and the standard genetic code was the best. With a 
measure which would be a good model for the errors against which the genetic code was optimized, the standard genetic code would 
then be found to be the global minimum code. There probably are no scientists who adhere to such an extreme variant of the 
"Error Minimization Hypothesis". It is, however, tempting to see the low $MS_0$ value as an indication that specific selection 
for error minimization was a major determinant of the codon assignments in the standard genetic code (e.g. \cite{wu}. \\
\\
\subsection{The Sequential "2-1-3" Model of Genetic Code Evolution}
\label{subsec:sequential}

Figure \ref{fig:samplesfinal3} shows the main result of \cite{freelandhurst98} remains valid when Space 0 is enlarged to
Space 1, and subsequently to Space 2: the MS value of the standard genetic code is \emph{better} than the MS value of the average
code when point mutations in the \emph{second} position are considered; it is \emph{much better} when point mutations in
the \emph{first} position are considered; and it is \emph{so much} better when point mutations in the \emph{third} position are 
considered that better codes in this respect are not visible in the graphs. This could point to the chronological order in
which the codon positions acquired coding information. Massey \cite{massey2006, massey2008, massey2010} published a series of 
papers in which the sequential acquisition of coding information by the second, then the first, and finally the third 
codon position is the major determinant of the codon assignments in the standard genetic code. According to this "2-1-3" 
model, the genetic code started with full degeneracy in the side positions. The amino acid repertoire would originally have 
been limited to four amino acids, and coding information was carried by the middle position. Subsequently the amino acid 
repertoire was expanded by assigning coding information to the first position. Because the code expansion would be "...facilitated 
by duplication of the genes encoding adaptor molecules and charging enzymes" \cite{massey2008}, amino acids of similar properties would be 
assigned to codons with the same middle nucleotide. Selection on error minimization plays a limited role in the "2-1-3" model in so far 
that code expansion via duplication of adaptor molecules followed by mutation of the middle position of the anticodon is selected 
against. Hence: "... amino acids of similar properties were selectively assigned to codons separated from one another by a single 
potential mutation" \cite{massey2006}. Finally, a further expansion was possible by assigning coding information to the third codon 
position. A consideration of the structure of the tRNA anticodon leads Massey to conclude that the third codon position is 
intrinsically the most error-prone. Therefore it is logical that distinguishing codons unambiguously on the third position is 
only possible when protein biochemistry has already progressed beyond the initial stages. Massey states that his analyses "...demonstrate 
that a substantial proportion of error minimization is likely to have arisen neutrally, simply as a consequence of code expansion, 
facilitated by duplication of the genes encoding adaptor molecules and charging enzymes. This implies that selection is at best 
only partly responsible for the property of error minimization" \cite{massey2008}. The concept of a genetic code in which 
coding information was carried by the middle position only, has been around since the sixties (e.g. with Crick: "For example, only 
the middle base of a triplet may have been recognized, a U in that position standing for any of a number of hydrophobic amino 
acids, an A for an acidic one etc." \cite{crick68}). The "2-1-3" model, however, goes further than that: it presents the chronological 
order in which the codon positions acquired coding information as the major determinant of the error minimization present in the code. 
The low $MS_0$ value is not incompatible with the "2-1-3" model; to the contrary, the "2-1-3" model is based on the low $MS_0$ value. \\
\\
\subsection{The Frozen Accident Theory}
\label{subsec:accident}

A third scenario is the Frozen Accident Theory of Crick \cite{crick68}. In this scenario, "... the actual allocation of amino acid to codons 
is mainly accidental and yet related amino acids would be expected to have related codons" \cite{crick68}. This is because there "...are 
several reasons why one might expect [...] a substitution of one amino acid for another to take place between structurally similar amino 
acids. First, [...] such a resemblance would diminish the bad effects of the initial substitution. Second, the new tRNA would probably start 
as a gene duplication of the existing tRNA for those codons. Moreover, the new activating enzyme might well be a modification of the existing 
activating enzyme. This again might be easier if the amino acids were related. Thus, the net effect of a whole series of such changes would 
be that \emph{similar amino acids would tend to have similar codons}, which is just what we observe in the present code" \cite{crick68}. Please 
note that in text preceding this fragment the possibility has been raised that "... the primitive tRNA was its own activating enzyme" \cite{crick68}, 
which is a description of a ribozyme \emph{avant la lettre}. At a certain moment the system would reach a stage in which "... more and more proteins 
would be coded and their design would become more sophisticated until eventually one would reach a point where no new amino acid could be introduced 
without disrupting too many proteins. At this stage the code would be frozen" \cite{crick68}. Please note that on the very first page of the 
paper the possibility is mentioned that the genetic code is not exactly identical for all organisms, although for widely different organisms 
it had been found to be very similar. Therefore the word "frozen" was probably from the start meant to be interpreted with a small degree of 
flexibility. The concept "relatedness" of amino acids is not rigorously defined in the paper, but Crick presents three examples of what he considers 
to be groups of related amino acids. "All codons with U in the second place code for hydrophobic amino acids". The polar requirements of this 
specific group of hydrophobic amino acids are 5.0, 4.9, again 4.9, 5.3, and 5.6. A second group of "related" amino acids is described in: "The basic 
and acidic amino acids are all grouped near together towards the bottom right-hand side ..." The polar requirements of this group of charged (and 
thus hydrophilic) amino acids are 10.1, 9.1, 13.0, and 12.5. The third example is the group of aromatic amino acids: "Phenylalanine, tyrosine and 
tryptophan all have codons starting with U". The polar requirements of these are 5.0, 5.4, and 5.2. Because "related" amino acids according to 
Crick tend to share a similar polar requirement, the low $MS_0$ value is not incompatible with the "Frozen Accident Theory". A clear difference 
between the "2-1-3" model and the "Frozen Accident Theory" is the presence of pairs of "related" amino acids with a second position difference 
in the latter: e.g. lysine-arginine, and phenylalanine-tyrosine. In this respect, it is relevant to observe that the MS value of the genetic 
code is lower than the MS value of the average code when point mutations in the second position are considered. Both the "2-1-3" model and the 
"Frozen Accident Theory" are scenarios in which the genetic code is basically a piece of historical information. Differences between these two 
scenarios are a lack of emphasis on sequential acquisition of coding information for the different codon positions in Crick's scenario; and a 
"refusal" by Crick to have a role for specific selection for error minimization in the scenario: "There is no reason to believe, however, that 
the present code is the best possible, and it could have easily reached its present form by a sequence of happy accidents. In other words, it 
may not be the result of trying all possible codes and selecting the best. Instead, it may be frozen at a local minimum which it has reached 
by a rather random path" \cite{crick68}. \\
\\
\subsection{The Stereochemical Theory}
\label{subsec:stereochemical}

A fourth scenario is what Crick named "The Stereochemical Theory" \cite{crick68}. According to this scenario there is a 
physico-chemical relationship between certain nucleic acid triplets and certain amino acids. The first such proposal was 
published by Gamow \cite{gamow}. Woese spent a lot of effort collecting evidence for the support of the Stereochemical
Theory \cite{woeseorder, woese65, woesebasis, woese66, woese67}. Orgel described this scenario as follows: "The simplest 
theory suggests that the role of tRNA's was originally filled by a set of much shorter polynucleotides, perhaps the 
anticodon trinucleotides themselves. In this form, the theory postulates that trinucleotides have a selective affinity 
for the amino acid coded by their complementary trinucleotide. Of course, the selectivity must have been limited in the 
first place, but it is argued that it might have been sufficient to produce primitive activating enzymes in the presence 
of a suitable messenger RNA. Then the system could have perfected itself by the "bootstrap" principle, [...]. If this 
type of theory is correct the code is not arbitrary; if life were to start again, certain features of the code would be 
reproduced because the physical interactions on which it is based are unchanging" \cite{orgel}. Exactly these kind of 
unchanging physical interactions are found in a number of recently published experiments (\cite{caporaso, yarusetal2005, yarusetal2009}
and references therein). Anticodons like GAA, GUA, GUG, and CCA are part of RNA molecules which bind respectively 
phenylalanine, tyrosine, histidine, and tryptophan. Again, phenylalanine and tyrosine form a group of amino acids 
coded by codons with U in the first position (contributing to a low $MS_0$ value), but in this scenario the formation 
of the group is due to a straightforward binding affinity of a GAA-containing RNA for phenylalanine, and another one 
of a GUA-containing RNA for tyrosine. Earlier experimental work pointed to a stereochemical relationship between the 
anticodons GCC, AGC and GAC and the simple amino acids glycine, alanine and valine respectively \cite{shimizu95}. The 
same author published models in which e.g. asparagine and lysine were shown binding their cognate anticodons \cite{shimizu}. 
If the major determinant for the codon assignments in the standard genetic code is stereochemical affinity between 
triplets and amino acids as reported in these publications, this implies a low $MS_0$ value. Therefore, the Stereochemical 
Theory is not incompatible with a low $MS_0$ value. \\
\\
\subsection{A four-column theory for the origin of the genetic code}
\label{subsec:column}

The four scenarios discussed above share the characteristic that one factor (either "minimization", "history" or 
"stereochemistry") is the major determinant of the codon assignments in the standard genetic code. They share this 
characteristic with the scenarios published by Wong \cite{wong75} and by Ikehara \cite{ikehara2002}. Other scenarios 
are present in which all three factors are major determinants  \cite{knightfreeland99, gulik}. As a last scenario, 
we discuss the four-column theory published by Higgs \cite{higgs2009}. Like the scenario proposed by Massey, the 
earliest genetic code according to the four-column theory is encoding a repertoire of four amino acids. Higgs 
is very detailed on the amino acids and the codon assignments in this earliest genetic code: the sixteen codons with U 
in the middle originally encoded valine, the sixteen middle-C codons alanine, the sixteen middle-A codons aspartate, 
and the sixteen middle-G codons glycine. Later amino acids were added to this code by a process of subdivision of 
these 16-codon blocks, in which a subset of the codons assigned to an early amino acid were reassigned to a later 
amino acid. In the four-column theory, codons with a certain middle position are reassigned to amino acids similar 
to the one originally assigned to codons with this middle position because this is the least disturbing to already 
existing protein sequences. The driving force for the reassignment is the "positive selection for the 
increased diversity and functionality of the proteins that can be made with a larger amino acid alphabet"\cite{higgs2009}.
An intermediate code is presented, with Leu, Ile and Val coded by middle-U codons, Ser, Pro, Thr and Ala coded by 
middle-C codons, Asp and Glu coded by middle-A codons, and all middle-G codons coding Gly. At this stage, the total 
of protein-coding sequences starts to influence the further development of the code strongly (code-message coevolution,
as in the series of papers by Sella and Ardell \cite{ardell2001, sella2002, ardell2002, sella2006}) because, as a 
consequence of their function in proteins, glycine codons become rare codons. The consequence of this is that the 
constraint to reassign them is relaxed. The final result is that amino acids which are not similar to glycine, but 
which are associated with strong positive selection because they bring radical new functions for proteins (cysteine, 
tryptophan and arginine) are found coded by middle-G codons. Although Higgs emphasizes that the driving force 
during the process of expansion of the amino acid repertoire is not the minimization of translational error, the 
four-column theory is not as "neutral" as the "2-1-3" model, because the "minimal disruption to the proteins already 
encoded by the earlier code" by adding "...later amino acids into positions formerly occupied by amino acids with 
similar properties" is such an important component of the scenario. \\
\\
Like the other discussed scenarios, the four-column theory is compatible with a low $MS_0$ value. All five discussed 
scenarios agree that error robustness due to codon assignments is present in the standard genetic code. The scenarios 
differ in the way they propose the error robustness has been built.

\subsection{Consequence of the error robustness}
\label{subsec:consequence}

The consequence of the error robustness is an enormous
potential to evolve. A variation in an RNA sequence can have different kinds of
consequences in the protein sequence. At the one end of the spectrum, the different
codon does not lead to a different amino acid. Slightly more effect would be that a
different codon would lead to a different amino acid, but this would be so similar to
the original amino acid that no difference in protein structure is the consequence.
Most important would be the effect that there is a difference in protein structure, but
so small that natural selection can use it as a slight step along an evolutionary path. At
the far end of the spectrum, finally, we find the lethal mutations. Because of this
graded intensity of evolutionary effect, the nature of the relationship between RNA
sequence and protein sequence (i.e. the genetic code) gives biochemistry an enormous
evolvability \cite{wagner, zhu}. This not necessarily implies that the phenomenon itself is built by
direct optimizing selection for the error minimizing aspects (exactly the same argument holds
for the aspects of stop codons allowing additional information to be encoded in protein-coding
sequences as described by Itzkovitz and Alon \cite{itzkovitzalon2007}).

\section{Conclusions}
\label{sec:conclusions}

Formulating the minimisation problem as a Quadratic Assignment Problem, we certify that 3.489, Goldman's
best solution \cite{goldman93} is in fact the optimal one. In spite of its theoretical hardness,
the size of the problem allows for exact solution methods instead of mere heuristics, that may fail
in finding the optimal solution. We demonstrated that it is possible to sample from much larger and more 
realistic code spaces. Leaving Space 0, and using simple wobble rules we constructed four progressively larger
code spaces. Their size is of a completely different order than that of Space 0. Spaces 3 and 4 contain all 
existing genetic code variations. Using a modified MS measure, the nature of Spaces 1 and 2 could be investigated. 
In Spaces 1 and 2, the standard genetic code was found to be a little more error robust when compared to randomly 
generated codes than it was found to be in Space 0. Finally, limitation of error robustness as a means to decide 
between different evolutionary scenarios is discussed.

\section{Materials and methods}
\label{sec:materials}

\subsection{Quadratic Assignment Problem}
\label{subsec:qap}

We formulate determining the minimum $MS_{0}$ as a Quadratic Assignment Problem. We use the graph
model presented in the Section~\ref{sec:background} for adjacency of the codon pairs.
We number the
amino acids $A_1,\ldots,A_{20}$ and the blocks in the standard genetic code $B_1,\ldots,B_{20}$. We
introduce binary decision variables $x_{ik}$, $i=1,\ldots,20$, $k=1,\ldots,20$; $x_{ik}$ gets value
$1$ if amino acid $A_i$ is assigned code block $B_k$ and value $0$ otherwise. If $x_{ik}=1$ and
$x_{j\ell}=1$ then this contributes to the objective a value
\begin{eqnarray*}
d_{ikj\ell}= \sum_{c\in B_k, c'\in B_l, \{ c,c'\} \in E} \left( r(A_i)-r(A_j) \right) ^2.
\end{eqnarray*}
To find the code with minimum $MS_0$-value we minimise
\begin{eqnarray*}
\sum_{i=1}^{20} \sum_{j=1}^{20} \sum_{k=1}^{20} \sum_{\ell=1}^{20}  d_{ikj\ell}x_{ik}x_{j\ell},
\end{eqnarray*}
subject to the restrictions
\begin{eqnarray*}
\sum_{i=1}^{20} x_{ik}=1,\ {\rm for} \ k=1,\ldots,20,
\end{eqnarray*}
\begin{eqnarray*}
\label{restrictionk}
\sum_{k=1}^{20} x_{ik}=1,\ {\rm for} \ i=1,\ldots,20,
\end{eqnarray*}
ensuring that each block encodes some amino acid and that each amino acid is encoded by some block, and
the restrictions
\begin{eqnarray*}
x_{ik}\in \{ 0,1\} ,\ {\rm for} \ i=1,\ldots,20,\ k=1,\ldots,20,
\end{eqnarray*}
ensuring that blocks cannot be assigned fractionally to some amino acid and
for another fraction to some other amino acid. \\

\noindent A similar model can be used to compute the code achieving minimum $MS_{0}^{S}$ value,
although it requires time in the order of weeks to compute, as opposed to hours for the $MS_{0}$
value. Further extending the above model, to compute the minima of the even larger code spaces,
leads to programs that even state-of-the-art algorithms cannot solve in any reasonable amount of
computer time.

\subsection{Counting and sampling}
\label{subsec:counting}

\begin{table}[h]
\caption{tRNA induced counts}
\centering
\begin{tabular}{|c|c|c|c|l|}
\hline
Amino acids & Stop codons & Unassigned & Multiplicity & tRNA Patterns\\
\hline
0& 0& 4& 1& ``uuuu''\\
\hline
0& 1& 3& 2& ``uusu'' ``uuus''\\
\hline
0& 2& 2& 1& ``uuss''\\
\hline
1& 0& 0& 1& ``aaaa''\\
\hline
1& 0& 1& 2& ``aaau'' ``aaua''\\
\hline
1& 0& 2& 2& ``aauu'' ``uuaa''\\
\hline
1& 0& 3& 1& ``uuua''\\
\hline
1& 1& 0& 2& ``aaas'' ``aasa''\\
\hline
1& 1& 1& 2& ``aasu'' ``aaus''\\
\hline
1& 1& 2& 1& ``uusa''\\
\hline
1& 2& 0& 1& ``aass''\\
\hline
2& 0& 0& 2& ``aaab'' ``aabb''\\
\hline
2& 0& 1& 1& ``aaub''\\
\hline
2& 1& 0& 1& ``aasb''\\
\hline
\end{tabular}
\label{tab:tab7}
\end{table}

\noindent
In Table \ref{tab:tab7} we have listed the possible ways to fill a single box that are compatible with the considered tRNA wobble rules.
Let $\{p_1, \ldots, p_{M}\}$ enumerate the possible tRNA patterns as listed in the
rightmost column of Table \ref{tab:tab7}. We write $a(p)$, $s(p)$, $u(p)$ for the number of amino acids, stop codons and unassigned codons present in pattern
$p$.\\
\\
\textbf{Problem.} We now consider the problem of filling 16 boxes (64 codons in total) using 20
different amino acids, $s$ stop codons and $u$ unassigned codons. It is useful to solve a slightly
more general problem: the number of ways to fill $b$ boxes using
\begin{itemize}
\item $N$ amino acids,
\item each of the first $a$ amino acids at least once,
\item exactly $s$ stop codons, and
\item exactly $u$ unassigned codons.
\end{itemize}
The original problem is obtained by setting $a=N=20$ and $b=16$.\\
\\
\textbf{Recurrence.} We denote the number of such fillings by $\#_N(b,a,s,u)$ and compute their
values by the recurrence
\begin{multline} \label{eq:main}
\#_N(b,a,s,u) = \\ \sum_{i=1}^M \sum_{j=0}^{a(p_i)} a(p_i)! \binom{a}{j} \binom{N-a}{a(p_i) - j} \#_N(b-1, a-j, s-s(p_i), u-u(p_i)).
\end{multline}
with basis
\begin{align*}
\#_N(0,0,0,0) &= 1,
\\
\#_N(b,a,s,u) &= 0 \quad\text{if $4b < a + s + u$},
\end{align*}
\textbf{Rationale. }The reasoning behind \eqref{eq:main} is the following. We fill box number $b$
first, and worry about the remaining boxes later. We iterate over the possible tRNA patterns with
variable $i$. To realise pattern $p_i$ we need $a(p_i)$ amino acids, $s(p_i)$ stop codons and
$u(p_i)$ unassigned codons. There is only one way to choose stop codons and unassigned codons, but
we can obtain the amino acids from two sources. We can take some from the $a$ still-to-use amino
acids that we have to use at least once, and we must take the others from the $N-a$ free amino
acids that can be used as desired. We consider all possible ways to realise the choice: We first
iterate over the number of amino acids that we take from the still-to-use pool with variable $j$.
Selecting $j$ out of $a$ still-to-use amino acids can be done in $\binom{a}{j}$ ways. Similarly,
taking the remaining $a(p_i)-j$ amino acids from $N-a$ free amino acids can be done in
$\binom{N-a}{a(p_i)-j}$ ways. All these $a(p_i)$ chosen amino acids are different, and so there are
$a(p_i)!$ ways to instantiate the pattern using them. Now we still have to fill the
remaining $b-1$ boxes, using the remaining $a-j$ still-to-use amino acids at least once, while using exactly $s-s(p_i)$ stop codons and leaving $u-u(p_i)$ codons unassigned.\\
\\
\textbf{Implementation. }The value $\#_{20}(16, 20, s, u)$ can be efficiently evaluated by dynamic programming. This is achieved by
storing all intermediate values of $\#$ that are computed in memory, and recalling them when they are needed instead of reevaluating $\#$. This way, $\#_N(b,a,s,u)$ can be evaluated in time
and space $O(b a s u)$. Note that a single call to $\#_N(b,a,s,u)$ computes $\#_N(b',a',s',u')$ for many $b' \le b$, $a' \le a$, $s' \le s$ and $u' \le u$.\\
\\
\textbf{Sampling. }The above dynamic programming implementation has the advantage that it allows uniform sampling over the space of all codes. We first sample a number uniformly between $1$
and $\#_N(b,a,s,u)$. Then we use the recurrence in reverse to determine which code this number corresponds to. This is done as follows. Say the number sampled was $n$. We then
incrementally evaluate the sum of \eqref{eq:main}. Once the partial sum up to $i$ surpasses $n$, we know that pattern $p_i$ was used in code number $n$.
Similarly we decode which amino acids are used and in which order they are placed. By explicitly keeping track of the set of still-to-use amino acids we can retrieve the entire code recursively.

\section{Acknowledgements}
We gratefully acknowledge Steven de Rooij for taking part in the effort to determine the global
minimum. Furthermore we gratefully acknowledge the large improvements of the manuscript by comments 
from Paul G. Higgs and two anonymous referees.

\end{document}